\documentclass[prd,aps,12pt]{revtex4}
\usepackage{epsf}
\def\be{\begin{equation}}
\def\ee{\end{equation}}
\def\ba{\begin{eqnarray}}
\def\ea{\end{eqnarray}}

\def\Re{\mbox{Re}}
\def\Im{\mbox{Im}}
\def\const{{\rm const.}}
\begin{document}

\title{Bulk black hole, escaping photons, and bounds on violations of Lorentz 
invariance}
\author{S. Khlebnikov}
\affiliation{Department of Physics, Purdue University, West Lafayette, IN 47907, 
USA}

\begin{abstract}
There are reasons (which we enumerate) to think that an infinite
extra dimension will harbor a black hole. In this case, brane-localized
modes of gravity and gauge fields become quasilocalized, and
light from a distant object can become extinct as it is lost to the black 
hole. In a concrete scenario, where the photon is localized
by gravity, we find that the extinction rate for propagating photons
is at least comparable to the correction to the real part of the frequency. 
That results, for example, in a stringent bound on renormalization of 
the speed of light.
\end{abstract}
\maketitle

\section{Introduction and results} \label{sect:intro}
In models with infinite extra dimensions
(for a review, see Ref. \cite{Rubakov:2001kp}), the four-dimensional nature 
of the observable world is due to the presence of special ``localized'' modes
for all the known particles and gravity. These modes are localized on
a submanifold (now referred to as ``brane'') in a higher-dimensional spacetime
(``bulk'') \cite{Joseph}. One
can consider localization by a scalar field \cite{Rubakov:1983bb}
(a mechanism useful for producing chiral fermions) 
or by gravity alone \cite{Visser:1985qm}. A localized mode for the graviton 
is known to exist in the case when the bulk is locally anti-de Sitter (AdS)
with a vanishing horizon \cite{Randall:1999vf}.

Backgrounds with non-vanishing horizons are also of interest. For reasons that
will be enumerated shortly, in the present paper, we consider 
geometries of the AdS-Schwarzschild type:
\be
ds^2 = - \frac{r^2}{R^2} \left( 1 - \frac{r_0^{d+1}}{r^{d+1}} \right)  dt^2 
+ \frac{dr^2}{\kappa^2 r^2 \left( 1 - \frac{r_0^{d+1}}{r^{d+1}} \right)}  
+ r^2 \sum_{i=1}^d (d\xi^i)^2 \equiv g_{MN} dx^M dx^N \; .
\label{ds2} 
\ee
There is a black-hole horizon at $r=r_0$ and a brane at $r=R$
(so, the extra dimension is strictly speaking not infinite, but rather 
``very large''); $\kappa$ is the inverse
AdS radius, and the coordinates $\xi^i$ span a $d$-dimensional torus. 
Only the region $r_0 < r \leq R$ will be important in what follows. 
In this region, the metric (\ref{ds2}) solves the $(d+2)$-dimensional vacuum 
Einstein equations with a negative cosmological constant:
\[
{\cal R}_{MN} = - (d+1) \kappa^2 g_{MN} \; .
\]
It is a particular case of the class of solutions described in 
Ref. \cite{Birmingham:1998nr}.

The coordinates $\xi^i$ are all periodic but with different periods. 
Three of them ($i=1,2,3$)
are periodic with period $2\pi$ and correspond to the three known spatial
dimensions (thus $R$ has the meaning of the ``size of the universe''; we take
$R\sim 1$ Gpc). The periods of the remaining $d-3$ of $\xi^i$ are much smaller,
so these are so far unobserved compact dimensions. Note that all $d$
dimensions are warped, with the same warp factor equal to $r^2$.

We consider the case when the parameter $R$ is time-independent, i.e., we do not
consider cosmology of the background (\ref{ds2}). 
It has been noted \cite{Chatillon:2006vw} that for $d > 3$ the simplest version 
of such a cosmology is problematic: if all dimensions (including 
the compact ones) expanded at the same rate, the fine structure constant would
be changing too fast. However, such a uniform expansion may be too
strong a condition to assume. Indeed, the sizes of the compact dimensions are
at this point arbitrary parameters (moduli of the solution). In
a more complete theory, they may be set to definite values by a weak potential.
We therefore consider the question of whether Eq. (\ref{ds2}) with $d > 3$
can be the late-time limit of a sensible cosmology as still open.

For $r_0=0$, the $d=3$ and $d=4$ versions of the metric (\ref{ds2})
are familiar from the studies of codimension-1 
(wall-like) \cite{Randall:1999vf} and codimension-2 (string-like) 
\cite{Gherghetta:2000qi} branes. (In this case, the metric is often written in terms 
of the radial coordinate $z$ related to $r$ by $r= R e^{-\kappa z}$.)
Our reasons for considering the case with the black hole ($r_0 > 0$) are as follows.

First, there is an argument \cite{Hebecker:2001nv} that a bulk black hole 
must have formed at some time 
during cosmological history. We see no reason why that black hole should have
completely evaporated by now. Second, when the number of spatial dimensions is large
enough, $d>3$, the $r_0 =0$ metric localizes gauge fields
\cite{Oda:2000zc,Dubovsky:2000av}. In this case, however, there is
a conical singularity at $r=0$. 
Ways of resolving the singularity, while preserving the 4-dimensional Lorentz 
invariance, have been proposed \cite{Ponton:2000gi}. On the other hand, perhaps the
simplest way to get rid of the singularity is to hide it behind a black-hole horizon, 
as achieved by the metric (\ref{ds2}) with $r_0 > 0$. 
Lorentz invariance is now broken,
but the strength of this breaking is controlled by the parameter $r_0/R$ and is
small if that parameter is small. Finally, in the extra-dimensional
solution to the strong CP problem 
\cite{Khlebnikov:1987zg,Khlebnikov:2004am,Khlebnikov:2006yq}, 
instanton transitions are viewed as transport of topological charge 
across the brane, with the result being ``recorded'' by the 
extra-dimensional physics. Topological charge falling into a bulk black
hole (a process presumably leading to an increase in the horizon size) seems 
an acceptable recording mechanism.

Now, one may consider cutting the horizon away---for example, by placing a second 
brane at some $r = r_- > r_0$. Indeed, for $d=3$ (one extra dimension), 
this has been a popular framework for 
Lorentz-invariance breaking phenomenology 
\cite{Kraus:1999it,Csaki:2000dm,Cline:2003xy}. Here, on the other hand, we consider
the entire region $r_0 < r \leq R$ and ask how the presence of the horizon
affects propagation of excitations on the brane. In particular,
we are interested in propagation of the transverse electromagnetic field 
$A_j$, where $j$ corresponds to one of the $\xi$ coordinates. For concreteness, 
we fix the boundary condition to be
\be
\left. F_{rj} \right|_{r=R} = (\partial_r A_j - \partial_j A_r)|_{r=R} = 0 \; .
\label{bc0}
\ee

Our main goal was to compute the decay rate of the localized (now,
quasilocalized) photon mode due to the photon
leaking into the black hole. That such a decay rate must appear is clear
from the asymptotic behavior of the mode functions near the horizon: at large 
values of the variable
\be
 x = - \ln \left( 1 -  \frac{r_0^{d+1}}{r^{d+1}} \right)
\label{x}
\ee
the mode functions corresponding to nonzero frequencies $\omega$ go as 
\be
f(x,t) \sim \exp\left\{ -i \omega t \pm i \frac{\omega R x}{(d+1) \kappa r_0}  
\right\} \; .
\label{plane}
\ee
For $\omega > 0$, both modes are regular, so only continuum stable states exist.
Similar considerations apply to a massless scalar (a mimic of the graviton). 

In the latter case (scalar), the absence of a normalizable mode for $r_0 > 0$
(and $d=3$) was noted in Ref. \cite{Kraus:1999it}. 
It was interpreted there as a consequence 
of the geometry becoming unreliable near $r=r_0$ for small $r_0$. However, 
Eq. (\ref{plane}) applies for {\em any} $r_0 > 0$, even those for which 
the surface gravity at the horizon is far from the Planck scale. For such $r_0$, 
we interpret Eq. (\ref{plane})
as a signal that the formerly discrete localized mode now becomes 
a resonance with a finite decay width into the continuum. Note that for massive
particles a similar effect occurs even in the absence of a black hole, i.e., 
for $r_0=0$ \cite{Dubovsky:2000am}.

We find that, for all but the smallest values of the momentum, the photon dispersion
law $\omega(k)$ has a sizable real part $\Re \omega(k) \approx \pm k$. We call this
the {\em propagating} regime.
More precisely, for the branch with $\Re \omega(k) > 0$, we obtain the following
results for the photon ``mass'' $m^2 = \omega^2 - k^2$ ($k$ is 
the 3-dimensional momentum):
\be
m^2(k) \approx \left\{ \begin{tabular}{lr} 
$-\epsilon k^2 + 1.019 e^{-2\pi i/3} [(d+1) \epsilon \kappa k^2]^{2/3} \; ,$  &
~~~~~$k \gg \kappa / \sqrt{\epsilon} \; ,$  \\
$ (d-1)(d-3) \kappa^2 e^{-i\pi \mu} \frac{|\Gamma(-\mu)|}{\Gamma(\mu)}
\left[ \frac{\sqrt{\epsilon} k}{(d+3) \kappa} \right]^{2\mu} \; ,$   &
~~~~~$\kappa \epsilon^{\frac{1}{d+1}} \ll k \ll \kappa / \sqrt{\epsilon} \; ,$  \\
$\const (-i k \kappa \epsilon^{\frac{d-2}{d+1}}) \; ,$   &
~~~~~$\delta \ll k \ll \kappa \epsilon^{\frac{1}{d+1}} \; ,$ 
\end{tabular}
\right.
\label{m2all}
\ee
where $\mu =\frac{d-1}{d+3}$, $\Gamma$ is Euler's gamma function, 
\be
\epsilon \equiv \left( \frac{r_0}{R} \right)^{d+1} 
\label{eps}
\ee
is the small parameter that measures the departure from Lorentz invariance, and 
$\delta$ is the boundary of the propagating regime (see Sect. \ref{sect:small} 
for details). (The branch with $\Re\omega < 0$,
has the opposite $\Re \omega$ but the same $\Im\omega$.)

The first two lines in (\ref{m2all}) are results of expansions in small parameters,
while the third line, in which ``\const'' is a positive numerical 
constant, is an order-of-magnitude estimate. The $-\epsilon k^2$ term in the first
line (the high-momentum regime) is a trivial renormalization of the speed of 
light into $v = (1-\epsilon)^{1/2}$,
due to the choice of units of length and time in Eq. (\ref{ds2}). 
[In the other two regimes, this term is subleading relative
to the terms included in Eq. (\ref{m2all}).] The high-momentum regime is
somewhat special: in it, the photon dissolves into a series of resonances whose
widths are of the same order as the distances between them; 
Eq. (\ref{m2all}) describes only the one with the smallest 
$|\Im \omega|$.

The main conclusion, then, is that, in the propagating regime, the photon decay rate
\[
\gamma(k) \equiv - 2 \Im\omega(k) \approx - \frac{\Im m^2}{k}
\]
is at least comparable to the correction to $\Re \omega$, i.e., to
$|\Re \omega(k) - v k|$.

Lorentz-noninvariant effects associated with $\Re \omega$, such as dependence
of the speed of light on momentum or difference between the limiting speeds 
of different particles, will be referred to as {\em kinematical}. In 
the present case, the {\em mere existence} of photons that reach 
us from distant sources imposes stringent bounds on such effects.
If photons with momentum $k$ reach us from a distance $l$, then there 
is a bound on $\gamma$: $\gamma(k) \alt l^{-1}$ and, 
in view of Eq. (\ref{m2all}), a related bound on $|d\Re \omega / dk  - v|$. 
Taking $l = 1$ Gpc, we obtain
\be
\left| \frac{d\Re \omega}{dk}(k)  - v \right| \alt 10^{-32}~\frac{\mbox{eV}}{k} \; .
\label{bound}
\ee
Note that this bound is independent of any constraints on the AdS parameter
$\kappa$, such as those
following from the experimental limits \cite{Hoyle:2004cw} on power-law corrections 
to Newton's law. Constraints on $\kappa$, however, are useful if we want to obtain 
a bound on the parameter $\epsilon$ itself. For example, consider
$d=5$ and use the second line in Eq. (\ref{m2all}). In this case, 
$\gamma = 2\kappa\sqrt{\epsilon}$, independently of $k$. Setting
$\gamma < (\mbox{1~Gpc})^{-1}$ and $\kappa > (\mbox{0.1~mm})^{-1}=2\mbox{~meV}$, 
we obtain $\epsilon(d=5) < 3\times 10^{-60}$. 

If a particle species (e.g., the electron) is tightly bound to the brane, 
the maximal propagation speed for it will be $v$. 
In this case, the bound (\ref{bound}) becomes a limit
on the difference between that maximal speed and the speed of light.
Such differences often lead to interesting effects \cite{Coleman:1998ti}, but in our
case, in view of the bound (\ref{bound}), they look prohibitively small.

A slightly better hope for detecting a bulk black hole may be offered by 
the extinction
effect itself. Indeed, as seen from Eq. (\ref{m2all}), in many cases $\gamma(k)$
grows with $k$, so one can imagine a situation when the apparent loss in the
luminosity of an object is negligible, say, in the optical part of spectrum but 
becomes significant for photons in the TeV range.

In the rest of the paper, after some preliminaries in Sect. \ref{sect:prelim},
we derive the three expressions presented in Eq. (\ref{m2all}) (Sects. 
\ref{sect:large}, \ref{sect:medium}, \ref{sect:small}). The corresponding 
expressions for a massless scalar are given in 
Sect. \ref{sect:scalar}. Sect. \ref{sect:concl} is a brief conclusion.

\section{Mode equation for the photon} \label{sect:prelim}
Equation for the electromagnetic field reads
\be
\partial_M [ \sqrt{-g} g^{MN} g^{PQ} F_{NQ} ] = 0 \; ,
\label{em}
\ee
where $g_{MN}$ is the metric extracted from Eq. (\ref{ds2}), 
$F_{MN} = \partial_M A_N - \partial_N A_M$, and the indices take values
$0,r$, or $i$, the latter running from 1 to $d$. We begin by fixing the gauge
$A_0 = 0$. Then, since the metric is static, the $P=0$ component of Eq. (\ref{em}) 
(the Gauss law) can be written as
\[
\partial_0 \partial_M [ \sqrt{-g} g^{MN} g^{00}  A_N ] = 0 \; ,
\]
which shows that, in addition to $A_0 = 0$, we can impose the ``Coulomb gauge''
condition
\be
\partial_M [ \sqrt{-g} g^{MN} g^{00}  A_N ] = 0 \; .
\label{coulomb}
\ee
Using this condition in the $P=r$ component of Eq. (\ref{em}), we obtain a closed 
equation for
$A_r$, which has the obvious solution $A_r = 0$. We concentrate on this type of
solutions in what follows.

The Coulomb gauge condition (\ref{coulomb}) is now simply
$\partial_i  A_i  = 0$.
Using this in the equation (\ref{em}) with $P=j$ and expanding $A_j$ in
Fourier components,
\[
A_j \sim \exp (-i\omega t + i R k^i \xi^i ) \; ,
\]
we bring the equation to the form
\be
\frac{g_{00}}{r^{d-2}} \partial_r [r^{d-2} g^{rr} \partial_r A_j ]
- k^2 \left( R^2 \frac{g_{00}}{r^2} + 1 \right) A_j = m^2 A_j \; ,
\label{m2}
\ee
where $m^2 = \omega^2 - k^2$. The values of $m^2$ determine the photon spectrum.

In terms of the variable $x$, defined by Eq. (\ref{x}), Eq. (\ref{m2}) becomes
\be
-(d+1)^2 \frac{\kappa^2 r_0^2}{R^2} (1- e^{-x})^{\frac{2d-2}{d+1}}
\partial_x [ (1- e^{-x})^{\frac{2}{d+1}} \partial_x A_j] 
- k^2 (1- e^{-x}) A_j =  m^2 A_j \; .
\label{m2x}
\ee
Eq. (\ref{x}) maps the range $r_0 < r \leq R$ to the range $x_0 \leq x < \infty$.
The boundary condition (\ref{bc0}) becomes
\be
\left. \partial_x A_j \right|_{x= x_0} = 0 \; .
\label{bc}
\ee
As we already mentioned, the ratio $r_0 / R$ must be small, to ensure the  smallness
of deviations from Lorentz invariance on the brane. As a result,
\[
x_0 = \epsilon + O(\epsilon^2)  \; ,
\]
where $\epsilon$ is the small parameter (\ref{eps}).

The ratio appearing
in front of the first term in Eq. (\ref{m2x}),
\be
(d+1) \frac{\kappa r_0}{R} \equiv 2 T \; ,
\label{T}
\ee
has a simple physical meaning: $T(1 - \epsilon)^{-1/2} \approx T$ is the 
temperature of the Hawking radiation from the black hole, as seen by an observer
on the brane. [This can be deduced by continuing the metric (\ref{ds2}) to the
Euclidean time $\tau = -i t$ and requiring that the period
of $\tau$ is such that the Euclidean geometry is smooth at the horizon---in 
the same way as the temperature was found for an AdS black 
hole with a spherical horizon in Ref. \cite{Hawking:1982dh}.]

We have not succeeded in solving Eq. (\ref{m2x}) exactly. So, in what follows
we consider limiting cases in which approximate expressions for $m^2$ can be
obtained.

\section{Escape near the brane} \label{sect:large}
We begin with the case when the photon momentum is large:
\be
\epsilon k^2 \gg \kappa^2 \; ,
\label{k-large}
\ee
where $\epsilon$ is the parameter (\ref{eps}).
In this case, the escape from the brane, i.e., the onset
of the oscillatory behavior of the modes, occurs at $x\approx x_0$, and we can 
approximate Eq. (\ref{m2x}) as follows:
\be
-4 T^2 x^{\frac{2d-2}{d+1}} \partial_x \left[ x^{\frac{2}{d+1}} 
 \partial_x A_j \right] - k^2 [\epsilon + (x-x_0)] A_j =  m^2  A_j \; .
\label{x-small}
\ee
Note that the second term on the left-hand side has been expanded near
$x=x_0$, while in the first term it is sufficient to expand near zero.

A change of variables, 
\be
x = \epsilon y^{d+1} \; ,
\label{y}
\ee
converts Eq. (\ref{x-small}) into
\be
- \kappa^2 y^{d-2} \partial_y \left[ y^{2-d} \partial_y A_j \right] 
- (d+1) \epsilon k^2 (y - y_0) A_j= (m^2 + \epsilon k^2) A_j \; .
\label{m2y0}
\ee
The range $x_0 \leq x < \infty$ is mapped to the
range $y_0 \leq y < \infty$ with $y_0 =1 + O(\epsilon)$.

Setting $A_j = y^{\frac{d-2}{2}} b$ brings Eq. (\ref{m2y0}) to the Schr\"{o}dinger 
form:
\be
-\kappa^2 \partial_y^2 b 
+ {1\over 4 y^2} d(d-2) \kappa^2 b - (d+1) \epsilon k^2 (y - y_0)  b = 
(m^2 + \epsilon k^2) b \; .
\label{sform}
\ee
To the accuracy indicated below, the ``centrifugal'' (second) term can be neglected,
and the equation becomes
\[
\kappa^2 \partial_y^2 b + (d+1) \epsilon k^2 (y-y_1) b  = 0 \; ,
\]
where
\[
y_1 = y_0 - \frac{m^2 + \epsilon k^2}{(d+1) \epsilon k^2} \; .
\]
Solutions are the Airy functions. In 
accordance with the general recipe for calculating resonance 
energies \cite{LL}, we pick the outgoing wave. For $\Re\omega > 0$, it is
\[
A_j(y) =  y^{\frac{d-2}{2}} 
\mbox{Ai} [ - \alpha^{1/3} e^{2\pi i /3} (y-y_1) ] \; ,
\]
where 
\[
\alpha = \frac{(d+1) \epsilon k^2}{\kappa^2} \; .
\]
Since $\mbox{Ai}(z)$ is an entire function, and $\alpha$ is a large parameter, 
the boundary condition (\ref{bc}) is satisfied near zeroes of
$\mbox{Ai}'(z)$. The first zero is at $z = a'_1 = -1.019$. Using that, 
we obtain
\be
m^2 = -  \epsilon k^2  + | a'_1|  e^{-2\pi i /3} 
[(d+1) \epsilon \kappa k^2]^{2/3} 
[1 + O(\alpha^{-1/3}) + O(\epsilon)] \; .
\label{m2high}
\ee
Other zeroes of $\mbox{Ai}'(z)$ correspond to resonances with larger $|\Im \omega|$.

\section{Escape at intermediate distances} \label{sect:medium}
Next, we consider cases when the inequality (\ref{k-large}) is reversed. 
In low dimensionalities, $d\leq 3$, the photon is delocalized even 
in the absence of a black hole (for $d=3$, this case was considered in Refs.
\cite{Davoudiasl:1999tf,Pomarol:1999ad,Bajc:1999mh}). We therefore concentrate 
on $d > 3$, when for $r_0 =0$ a localized mode exists 
\cite{Oda:2000zc,Dubovsky:2000av}. For $r_0 > 0$, however, photons can escape
(fall into the black hole). As we will now see, for photons with momenta in the
range
\be
T \ll   k \ll  \frac{\kappa}{\sqrt{\epsilon} } \; ,
\label{k-medium}
\ee
the escape distance $y_e$ falls in the range
\be
1 \ll y_e \ll \frac{R}{r_0} \; .
\label{y-medium}
\ee
For $y \ll R/r_0$, $x$ is still small, and we can approximate Eq. (\ref{m2x}) as 
\be
- \kappa^2 y^{d-2} \partial_y \left[ y^{2-d} \partial_y A_j \right] 
- \epsilon k^2 y^{d+1} A_j= m^2  A_j \; .
\label{m2y}
\ee
[Additional 
corrections---those from the higher powers of $x$ in the derivative term in
Eq. (\ref{m2x}) turn out to be negligible.]

At sufficiently small $y$, the $k^2$ term in (\ref{m2y}) can be neglected, and
an approximate solution can be obtained by expansion in $y$. To the required
accuracy,
\be
A_j \approx 1 +  c y^{d-1} + \frac{m^2 y^2}{2(d-3) \kappa^2}  \; ,
\label{sol-left}
\ee
where $c$ is an integration constant. It is fixed by the boundary condition 
(\ref{bc}):
\[
c = - \frac{m^2 y_0^{3-d}}{(d-1)(d-3) \kappa^2} \; .
\]
Note that, without the $k^2$ term, Eq. (\ref{m2y}) is precisely the mode equation
in the absence of a black hole, and indeed Eq. (\ref{sol-left}) can alternatively
be obtained from the exact solution found for that case in 
Ref. \cite{Dubovsky:2000av}.

On the other hand, for $y^{d+1} \gg |m^2|/ \epsilon k^2$, we can drop the $m^2$ 
term in Eq. (\ref{m2y}). Then, a change of variables,
\[
\zeta = \frac{2 \sqrt{\epsilon} k}{(d+3) \kappa} y^{d+3 \over 2} \; ,
\]
reduces Eq. (\ref{m2y}) to the Bessel equation of order 
\[
\mu = \frac{d-1}{d+3} \; .
\]
The outgoing wave (for $\Re\omega > 0$) is
\be
A_j(\zeta) = C \left( \zeta \over 2 \right)^\mu H^{(1)}_\mu(\zeta) \; ,
\label{sol-right}
\ee
where $H^{(1)}$ is the Hankel function, and $C$ is a constant. 

Oscillations of $H^{(1)}$ set in at $\zeta \sim 1$ or, equivalently, at
\be
y\sim  (\kappa / \sqrt{\epsilon} k)^{2\over d+3} \equiv y_e \gg 1 \; .
\label{y1}
\ee
At smaller $y$, we can use the small-argument expansion
\be
H_\mu^{(1)}(\zeta) \approx - \frac{i}{\pi}
\left\{  \Gamma(\mu) \left( \zeta \over 2 \right)^{-\mu}
+ e^{-i\pi \mu} \Gamma(-\mu) \left( \zeta \over 2 \right)^\mu 
\right\} \; .
\label{H}
\ee
We see that the
two terms in Eq. (\ref{H}) correspond to the first two terms in (\ref{sol-left}).
Therefore, to the leading order in the small
parameters, $C= i\pi / \Gamma(\mu)$, and
\be
m^2 = - (d-1)(d-3) \kappa^2 e^{-i\pi \mu} \frac{\Gamma(-\mu)}{\Gamma(\mu)}
\left[ \frac{\sqrt{\epsilon} k}{(d+3) \kappa} \right]^{2\mu} \; .
\label{m2medium}
\ee
Note that, since $0 < \mu < 1$, $\Gamma(-\mu)$ is negative.

For the above solution to be consistent, the escape distance (\ref{y1}) must be
much smaller than $R/r_0$, where the small-$x$ approximation breaks down. This
leads to the left inequality in (\ref{k-medium}).

\section{Escape near the horizon} \label{sect:small}
Photons with 
\be
k \ll T 
\label{k-small}
\ee
escape (for $d > 3$) at $x \gg 1$. 
We do not have an approximate solution that would allow us to
traverse the region $x\sim 1$ and so, for this case, limit ourselves to
an order-of-magnitude estimate of $m^2$.

Using Eq. (\ref{sol-left}) for the region $x \ll 1$ and the outgoing wave from
(\ref{plane}) for $x \gg 1$, and matching their logarithmic derivatives at 
$x=1$, we obtain 
\be
m^2 = \omega^2 - k^2 \sim - i \omega \kappa \left( \frac{r_0}{R} \right)^{d-2} \; .
\label{m2low}
\ee
Two limits of this expression are of interest. For
\[
\kappa \left( \frac{r_0}{R} \right)^{d-2} \ll k \ll T \; ,
\]
the photon is oscillating with a $k$-independent decay rate:
\[
\omega \pm k \sim -i \kappa \left( \frac{r_0}{R} \right)^{d-2} \; .
\]
In the opposite limit, 
\[
k \ll \kappa \left( \frac{r_0}{R} \right)^{d-2} \equiv \delta \; ,
\]
there is a mode with $\omega \sim -i \delta$, and another one with a curious 
diffusive behavior:
\[
\omega \sim -i \frac{k^2}{\delta}  \; .
\]
Thus, $k\sim \delta$ is the upper limit on the momentum of photons that can
propagate on the brane.

\section{Escape of a massless scalar} \label{sect:scalar}
By the same transformations as those used in Sect. \ref{sect:prelim}, the equation
\[
\partial_A (\sqrt{-g} g^{AB} \partial_B \phi) = 0
\]
can be brought to the form
\[
\frac{g_{00}}{r^d} \partial_r [r^d g^{rr} \partial_r \phi ]
- k^2 \left( R^2 \frac{g_{00}}{r^2} + 1 \right) \phi = m^2 \phi \; ,
\]
which differs from Eq. (\ref{m2}) for the photon only by the power in which
$r$ appears in the derivative term ($d$ instead of $d-2$). In terms of the
variable $x$,
\be
-(d+1)^2 \frac{\kappa^2 r_0^2}{R^2} (1- e^{-x})^{\frac{2d}{d+1}}
\partial_x^2 \phi
- k^2 (1- e^{-x}) \phi =  m^2 \phi \; .
\label{m2scalar}
\ee
We use the same boundary condition as for the vector:
\[
\left. \partial_x \phi \right|_{x= x_0} = 0 \; .
\]

The same three cases as those in Sects. \ref{sect:large}, \ref{sect:medium}, and
\ref{sect:small} can be considered. In the high-momentum case, 
$\epsilon k^2 \gg \kappa^2$, the transformation
$\phi = y^{\frac{d}{2}} \chi$ results in a Schr\"{o}dinger equation that
differs from Eq. (\ref{sform}) of Sect. \ref{sect:large} only by the coefficient 
of the ``centrifugal''term. Since that term has been dropped there anyway,
the result (\ref{m2high}) is unchanged.

For intermediate momenta, $T \ll   k \ll  \kappa/\sqrt{\epsilon}$, the method 
of Sect. \ref{sect:medium} applies. The counterpart of Eq. (\ref{m2y}) is
\be
- \kappa^2 y^d \partial_y \left[ y^{-d} \partial_y \phi \right] 
- \epsilon k^2 y^{d+1} \phi = m^2 \phi \; .
\label{m2y-scalar}
\ee
The scalar is quasilocalized for any $d > 1$, a condition we now 
assume is satisfied. The asymptotic form (\ref{sol-left}) is replaced 
with
\be
\phi \approx 1 + c' y^{d+1} + \frac{m^2 y^2}{2(d-1) \kappa^2} \; ,
\label{left-scalar}
\ee
where $c'$ is fixed by the boundary condition.
At $y^{d+1} \gg |m^2|/\epsilon k^2$, the equation is again approximately Bessel
but now of order 
\[
\mu' = \frac{d+1}{d+3} \; .
\]
Proceeding as in Sect. \ref{sect:medium}, we obtain
\be
m^2 \approx - (d^2 - 1) \kappa^2 e^{-i\pi\mu'} \frac{\Gamma(-\mu')}{\Gamma(\mu')}
\left[ \frac{\sqrt{\epsilon} k}{(d+3) \kappa} \right]^{2\mu'} \; .
\label{medium-scalar}
\ee
Finally, for $k \ll T$, Eq. (\ref{m2low}) is replaced with
\be
m^2 \sim - i \omega \kappa \left( \frac{r_0}{R} \right)^d \; .
\label{low-scalar}
\ee
The higher powers of $\epsilon$ in Eqs. (\ref{medium-scalar}) and (\ref{low-scalar}),
compared to those in Eqs. (\ref{m2medium}) and (\ref{m2low}), show that the scalar
is bound to the brane tighter than the photon.

\section{Conclusion} \label{sect:concl}
Our main (and admittedly somewhat pessimistic) conclusion is that, if our world
looks like the construction described in this paper, Lorentz invariance in 
it is very well protected, at least at present. Indeed, in this case
the rate at which a propagating photon ``decays'' 
(i.e., leaks into the black hole) is at least 
of the same order as the correction to the real part of the frequency.
Then, the very fact that photons can propagate over astronomical
distances imposes stringent bounds on kinematical violations of 
Lorentz invariance, such as dependence
of the speed of light on momentum or (under mild further assumptions) difference
between the limiting speeds of different particles.

In these circumstances, one may be compelled to look directly for traces of 
the extinction effect in astrophysical data.
That, however, would seem to require a rather detailed understanding of 
the intrinsic luminosity of individual sources over a broad range of 
photon frequencies.

Finally, if the black hole has electric charge or angular momentum, these
will lead to additional Lorentz-noninvariant effects, which may deserve
a further investigation.

The author thanks John Finley and Peter Tinyakov for discussions of the results.
This work was supported in part by the U.S. Department of Energy through Grant
DE-FG02-91ER40681 (Task B).

\end{document}